\documentclass[12pt,a4paper]{article}

\usepackage{arxiv}
\usepackage{xfrac}
\usepackage{graphicx}
\usepackage{amsmath,amssymb,amsfonts}
\usepackage{subcaption}
\usepackage{makecell}
\usepackage{xcolor}
\usepackage{textcomp}
\usepackage[numbers, sort&compress]{natbib}

\onecolumn

\makeatletter
\def\ps@pprintTitle{%
  \fancyhf{}%
  \renewcommand{\headrulewidth}{0pt}%
  \renewcommand{\footrulewidth}{0pt}%
  \fancyhead[C]{}%
  \fancyhead[R]{}%
  \fancyhead[L]{}%
  \fancyfoot[C]{\thepage}%
  \pagestyle{plain}%
}
\makeatother

\AtBeginDocument{%
  \onecolumn
  \pagestyle{plain}%
}

\usepackage[utf8]{inputenc}
\usepackage[T1]{fontenc}
\usepackage{hyperref}
\usepackage{url}

\title{Defining the Magnetization State of LCF Magnets: From Material Properties to Motor-Level Metrics}

\author{
  Taha El Hajji \\
  \textit{Alvier Mechatronics AB} \\
  Helsingborg, Sweden \\
  0000-0002-0168-8180 \\
  \And
  Aleksandr Nadkin \\
  \textit{Alvier Mechatronics AB} \\
  Helsingborg, Sweden \\
  0009-0004-3756-3103
  \And
  Stefan Skoog \\
  \textit{Alvier Mechatronics AB} \\
  Helsingborg, Sweden \\
  0000-0001-7697-5259
  \AND
  Lars Sjöberg \\
  \textit{Alvier Mechatronics AB} \\
  Helsingborg, Sweden
  \And
  Kristoffer Nilsson \\
  \textit{Alvier Mechatronics AB} \\
  Helsingborg, Sweden
  \And
  Anthony C. Morcos \\
  \textit{ACM Magnetics, Inc.} \\
  \textit{Raleigh, NC, United States}
}

\begin{document}
\let\preprintfooter\relax
\maketitle

\begin{abstract}
Variable flux memory motors, which employ Low Coercive Force (LCF) magnets, achieve extended high-efficiency operation through controllable magnetization states. To address the need for a unified approach to defining and comparing the magnetization state (MS) across material and motor levels, this paper proposes four MS definitions: two based on intrinsic material properties—magnetic flux density B and magnetic polarization J—and two based on motor-level quantities—fundamental flux linkage and back-EMF components. These definitions are evaluated across the id, iq operating plane using finite element analysis on an interior PMSM with a hybrid magnet configuration (LCF and HCF: High Coercive Force) and a defined circuit setup. The results clarify the relationship between material-level behavior and measurable motor quantities. The proposed framework provides guidance for selecting appropriate MS metrics depending on the application objective, whether for material analysis, control implementation, or condition monitoring in variable flux machines.
\end{abstract}


\section{Introduction}

Permanent magnet synchronous machines (PMSMs) dominate automotive and industrial applications due to their high torque density and efficiency \cite{ElHajji2023}. Most employ High Coercive Force (HCF) magnets like sintered NdFeB or SmCo, with intrinsic coercivities exceeding 1000 kA/m at room temperature, ensuring a constant magnetization state under normal operation. In contrast, Low Coercive Force (LCF) magnets e.g., FeN \cite{Niron2025}, AlNiCo, MnBi, and MnAl, also known as semi-hard magnetic materials, can be used in variable flux motors—often termed memory motors \cite{Jayarajan2019,Hua2019,Basnet2021,Owen2011,Zhou2022,LAbbate2023,Arribas2024}. These possess moderate intrinsic coercivity (100–300 kA/m at 20°C), enabling the magnetization level, and thus motor flux, to be intentionally varied via controlled current pulses. This capability extends the high-efficiency operating region \cite{Hsieh2022,Fukushige2015,Do2022,Liu2021,Huynh2023}.

Temperature further complicates matters. For most magnetic materials, rising temperature reduces intrinsic coercivity, shifting the demagnetization curve knee toward the origin \cite{demag,Ding2023}. Conversely, cooling increases coercivity, making intentional state changes more difficult. Some materials like MnBi exhibit a positive temperature coefficient of coercivity, offering improved high-temperature stability, while emerging materials like Iron Nitride (FeN) are nearly temperature-insensitive.

The concept of magnetization state (MS) in variable flux machines has been explored extensively \cite{Jayarajan2019,Jang2021}. Foundational work established that permanent magnet characterization relies on remanence, coercivity, and recoil permeability. Researchers have sought practical methods to define and estimate instantaneous MS. At the motor level, MS has been assessed through indirect metrics such as induced voltage and flux linkage. Applications demonstrate that MS changes can be observed through induced voltage variations, with demagnetization reducing voltage contribution from variable magnets while fixed-flux magnets maintain baseline flux. Control-oriented approaches using back EMF measurements have been developed for real-time MS estimation and closed-loop control. Advanced characterization techniques have also been developed to observe magnetization behavior at multiple scales.

Despite these advances, a systematic comparison of MS definitions spanning from fundamental material properties to measurable motor-level quantities is required. Furthermore, the relationship between these definitions across the full \(i_d, i_q\) operating plane has not been thoroughly investigated. This paper addresses this gap by proposing four definitions for the magnetization state of LCF magnets: intrinsic material properties based on magnetic flux density (\(\vec{B}\)) and magnetic polarization (\(\vec{J}\)), and motor-level metrics based on fundamental flux linkage and back-EMF components. These definitions are evaluated across the \(i_d, i_q\) plane using finite element analysis on a representative interior PMSM with hybrid magnet configuration.

This paper is organized as follows. Section 2 presents the nomenclature. Section 3 provides a background of magnets with a focus on LCF magnets. Section 4 describes the studied motor topology and simulation setup. Section 5 introduces the four proposed MS definitions. Section 6 presents the results comparing the definitions.

\section{Nomenclature}

\begin{tabbing}
\hspace{3.5cm} \= \kill
LCF   \> Low Coercive Force \\
HCF   \> High Coercive Force \\
MS    \> Magnetization State \\
PMSM  \> Permanent Magnet Synchronous Machine \\
MMF   \> Magnetomotive Force \\
EMF   \> Electromotive Force \\
FE    \> Finite Element \\
$B$          \> Magnetic flux density (T) \\
$J$          \> Magnetic polarization, $J = B - \mu_0 H$ (T) \\
$H$          \> Magnetic field strength (A/m) \\
$B_r$        \> Remanence (remanent flux density) (T) \\
$J_r$        \> Remanent polarization (T) \\
$iH_c$       \> Intrinsic coercivity (A/m) \\
$\mu_{\text{rec}}$ \> Recoil permeability (-) \\
$\mu_g$      \> Drooping permeability (-) \\
$R$          \> Round radius – knee curvature (A/m) \\
$PC$         \> Permeance coefficient \\
$i_d$, $i_q$ \> $d$-axis and $q$-axis currents (A) \\
$N$          \> Effective number of turns per pole \\
$l_m$, $A_m$ \> Magnet length (m) and cross-sectional area (m²) \\
$l_g$, $A_g$ \> Air gap length (m) and cross-sectional area (m²) \\
$\mu_0$      \> Permeability of free space (H/m) \\
$\Phi_{\text{fund}}$ \> Fundamental component of phase flux linkage (Wb) \\
$E_{\text{fund}}$    \> Fundamental component of back EMF (V)
\end{tabbing}

\section{Variable Magnets: Background}

This section establishes the foundational knowledge required to understand the behavior of Low Coercive Force (LCF) magnets and their application in variable flux motors. We begin by distinguishing LCF from traditional high coercivity magnets (e.g. NdFeB). The concept of the magnet operating point and its manipulation via stator current is explained using the load line approach. Finally, we review the magnetization and demagnetization mechanisms, and provide an overview of common LCF magnet types.

\subsection{Low Coercive Force (LCF) and High Coercive Force (HCF) Magnets}

Permanent magnets used in electrical motors can be broadly categorized by their ability to resist demagnetization, a property quantified by intrinsic coercivity. High Coercive Force (HCF) magnets, such as sintered NdFeB and SmCo, possess very high intrinsic coercivity (typically 1000 kA/m at 20C temperature). This ensures their magnetization state remains virtually constant under the influence of stator currents up to the motor's rated limits, making them ideal for fixed-flux motors where the air-gap flux is constant.

In contrast, Low Coercive Force (LCF) magnets are characterized by a moderate intrinsic coercivity, typically in the range of 100-300 kA/m at 20C temperature. This property places them in a unique category often referred to as semi-hard magnetic materials. This moderate coercivity is a feature that enables the magnet's magnetization level, and thus the motor's flux, to be intentionally and efficiently varied by applying controlled current pulses. This principle is the foundation of variable flux motors, also known as memory motors.

\subsection{Key Magnetic Properties}

\begin{figure}[htbp]
\centering
\includegraphics[width=0.7\textwidth]{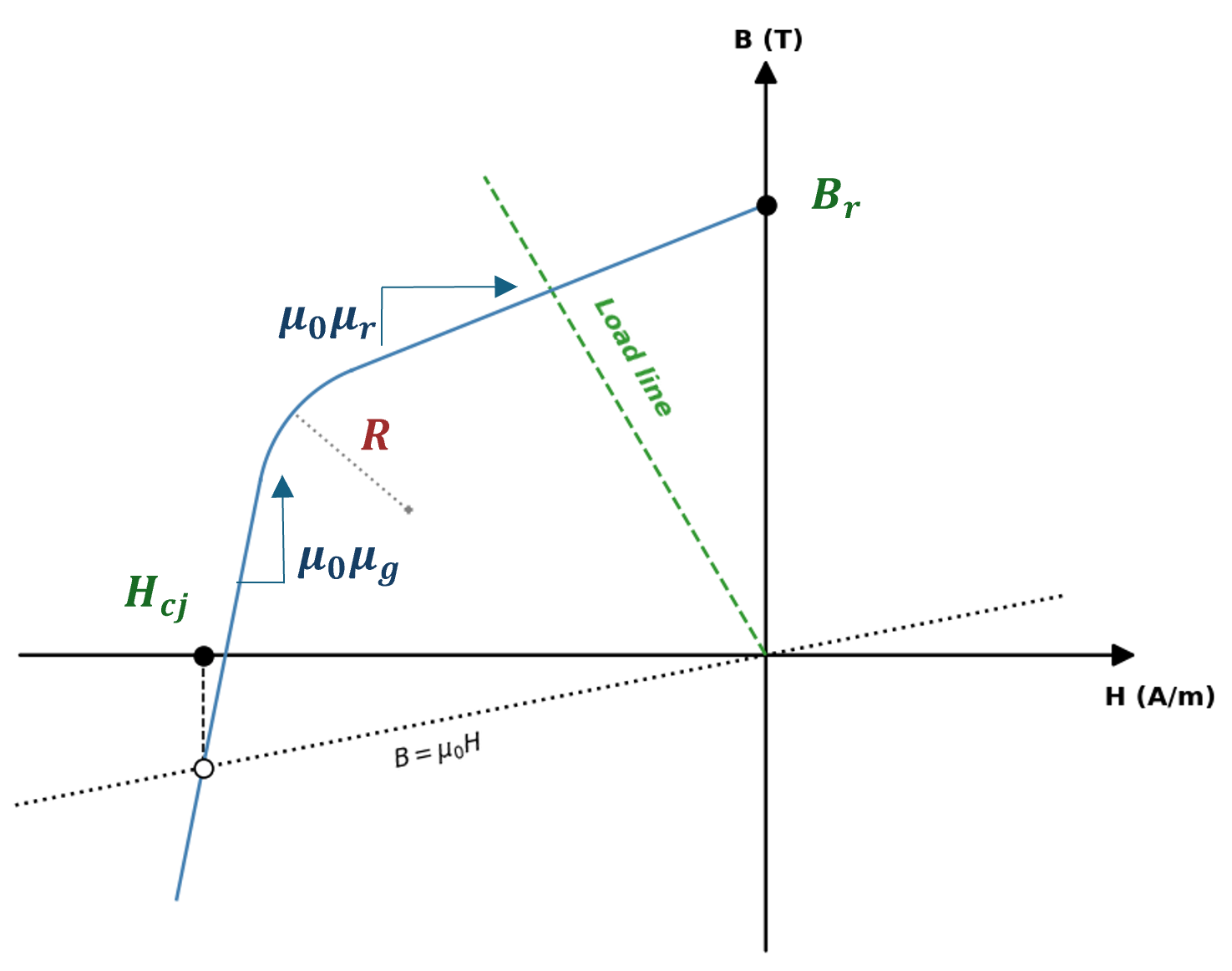}
\caption{A typical demagnetization BH curve in 2nd Quadrant for a magnet with key properties}
\label{fig1}
\end{figure}

The behavior of both LCF and HCF magnets is best understood through their B-H hysteresis loop characterized by the following key properties (Fig.~\ref{fig1}):

\begin{itemize}
    \item \textbf{Remanence ($B_r$)}: This is the residual magnetic flux density remaining in the material after it has been fully magnetized by a large external field and that field is subsequently reduced to zero.
    \item \textbf{Intrinsic Coercivity ($iH_c$)}: This is the magnitude of the demagnetizing field required to reduce intrinsic polarization ($J = B - \mu_0 H$) to zero. It is a fundamental measure of the resistance of a material to demagnetization.
    \item \textbf{Recoil Permeability ($\mu_{rec}$)}: During demagnetization and magnetization, the operating point returns along a "recoil line" belonging to a minor loop. The recoil permeability is the average slope of this recoil line ($\mu_{rec} \approx \Delta B / \Delta H$) which is often assumed to be constant.
    \item \textbf{Drooping Permeability ($\mu_{g}$)}: In the region around this knee, the incremental permeability decreases significantly. The variation follows a linear behavior with a slope of $\mu_{rec}$. This parameter is high for LCF magnets and low for HCF magnets.
    \item \textbf{Round Radius ($R$)}: A measure of the curvature at the knee of the demagnetization curve, representing the transition region where irreversible demagnetization begins.
\end{itemize}

\subsection{Magnet Operating Point and the Load Line}

The operating point of a magnet in a magnetic circuit is determined by the intersection of its material properties with the load line representing the characteristics of the external circuit. The latter represents the reluctance of the external magnetic circuit, including air gaps, iron paths, and any magnetomotive force (MMF) sources. In open circuit (i.e. no other MMF in the magnetic circuit), the intersection occurs either in the 2nd quadrant or the 4th quadrant. It is quantitatively defined by the permeance coefficient ($PC$) and the external MMF., which relates the magnet's flux density ($B_m$) and internal field ($H_m$) at the operating point.

For a given magnetic circuit geometry, the load line equation is derived from Ampere's law and flux conservation. Below the case of open circuit and the case with demagnetization current are presented:

\begin{itemize}
    \item \textbf{Case 1: Open Circuit}
\end{itemize}

In the scenario of magnets placed in the air, the magnet is subjected only to its own self-demagnetizing field arising from the geometry. The load line has the following equation:
\begin{equation}
B_m = -\mu_0 \times \left(\frac{l_m}{l_g} \frac{A_g}{A_m} \right) \times H_m = -\mu_0 \times PC \times H_m
\label{eq:loadline_base}
\end{equation}

where $l_m$ and $A_m$ are the length and the area of the magnet, respectively; $l_g$ and $A_g$ are the length and the area of the air gap, respectively; $B_m$ and $H_m$ are the flux density and the field strength in the magnet; $\mu_0$ is the permeability of air; and $PC=(l_m A_g)/(l_g A_m)$ is the permeance coefficient. The operating point is the intersection of this load line with the magnet's minor/major loops (Fig. \ref{fig1}).

For LCF magnets, the intersection of this load line with the magnet's BH curve occurs often below the knee region, which explains the natural demagnetization of magnets. For HCF magnets, low permeance coefficients, dictated by the aspect ratio, leads to irreversible demagnetization as well.

The equation \eqref{eq:loadline_base} applies also to magnets in motors without id current (assuming no cross coupling between q and d axis). The intersection of the load line with the magnet's BH curve establishes the motor's no-load operating point, as shown in Fig. \ref{fig1}. For a well-designed motor, this point is typically located on the linear region of the curve, above the knee, ensuring a stable flux level.

\begin{itemize}
    \item \textbf{Case 2: With $i_d$ Current}  
\end{itemize}
When a d-axis current ($i_d$) is applied, the resulting armature MMF acts directly on the magnetic circuit. This MMF is equivalent to an additional magnetomotive force that effectively shifts the load line horizontally. Ampere's law now includes the armature MMF: $H_m l_m + H_g l_g = N i_d$, where $N$ is the effective number of turns per pole. Solving this together with the flux conservation equation yields the modified load line expression:

\begin{equation}
B_m = -{\mu_0} \times PC \left( H_m - N i_d / l_m \right)
\label{eq:loadline_current}
\end{equation}

\begin{figure}[htbp]
\centering
\includegraphics[width=0.6\textwidth]{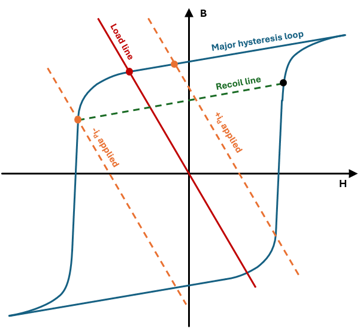}
\caption{Load line in case of open circuit and d-axis current}
\label{fig2}
\end{figure}

During demagnetization, we have a negative $i_d$, shifting the load line leftward by a constant horizontal offset. The new operating point is found at the intersection of this shifted load line with the minor/major loops, and it will be at a lower flux density than the no-load point, as illustrated in Fig. \ref{fig2}. Inversely, during magnetization, we apply a positive $i_d$, shifting the load line rightward by a constant horizontal offset. This leads to a new operating point with higher flux density than the no-load point, as illustrated in Fig. \ref{fig2}.

The critical consequence occurs if this new intersection falls below the knee region of the minor/major loop. At this point, the magnet experiences irreversible demagnetization. When the current pulse ($i_d$) is removed, the load line shifts back to its original, geometry-defined position described by \eqref{eq:loadline_base}. However, the magnet's operating point does not return to the original no-load point. Instead, it moves along a new path—a recoil line with a slope of approximately $\mu_{rec}$ to a new operating point on the recoil line, corresponding to a permanently reduced remanent flux. It is worth mentioning that the load line formulation represented here is simplified and does not consider the impact of q-axis current on demagnetization, therefore no cross-coupling between q-axis and d-axis is considered.

\subsection{Magnetization and Demagnetization Mechanisms}

The ability to control the magnetization state of an LCF magnet is fundamentally a controlled manipulation of its recoil behavior. Starting from a fully magnetized state at the remanence point ($B_r$) on the major hysteresis loop, applying an external demagnetizing field ($H_{demag}$) drives the operating point down along the major demagnetization curve. If this field is removed before the operating point reaches the knee region, the material will retrace its path back to $B_r$ along the major loop, effectively behaving as a reversible magnet with permeability $\mu_{rec}$. This is a condition of no net change in magnetization.

\begin{figure}[htbp]
\centering
\includegraphics[width=0.5\textwidth]{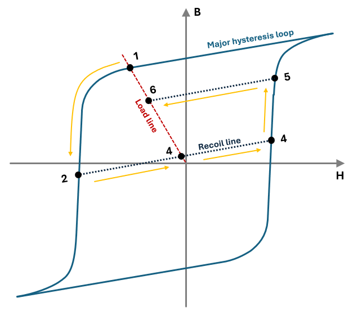}
\caption{Recoil lines and minor/major loops}
\label{fig3}
\end{figure}

To alter the state, the demagnetizing field must be sufficiently strong to push the operating point past the knee and onto the non-linear portion of the curve. When the field is removed from this new point, the operating point no longer returns to $B_r$. Instead, it follows a recoil line with a slope of approximately $\mu_{rec}$ to a new remanent flux density, $B_r'$, which is lower than the original $B_r$, as shown in Fig. \ref{fig3}. The magnet is now in a partially demagnetized state. The process of remagnetization is symmetric: applying a strong magnetizing field will drive the operating point up following the recoil line until reaching the major loop. Once the applied external magnetizing field is removed, the operating point follows a new recoil line from (crossing the last achieved flux density) until it meets the load line in open circuit case, leading therefore to a higher $B_r$, as shown in Fig. \ref{fig3}.

Temperature plays a critical role in these magnetization and demagnetization processes. For most magnetic materials, an increase in temperature causes a reduction in intrinsic coercivity, effectively shifting the knee of the demagnetization curve toward the origin. This means that a magnet which is fully stable at room temperature may become susceptible to irreversible demagnetization at elevated operating temperatures. Unlike most LCF magnets, Iron Nitride magnets (FeN) manufactured by Niron provide the property of being relatively insensitive to temperature.

\subsection{Examples of LCF Magnet Materials}

\begin{figure}[htbp]
\centering
\includegraphics[width=0.6\textwidth]{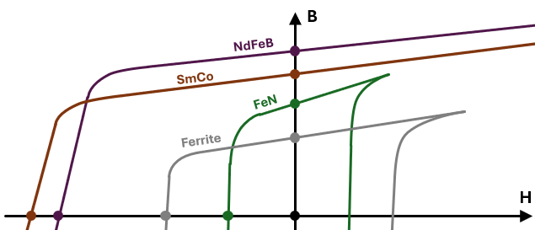}
\caption{BH curve of magnets}
\label{bh_comparison}
\end{figure}

Several families of magnetic materials exhibit the semi-hard properties required for variable flux applications. Table~\ref{tab:materials} summarizes the key properties of common LCF magnet materials alongside a typical NdFeB magnet for reference. Ranges are indicative and may vary with specific composition and manufacturing process. Figure \ref{bh_comparison} illustrates typical demagnetization curves for LCF magnets alongside a high-coercivity NdFeB magnet at $20^\circ\text{C}$ temperature.

\renewcommand{\arraystretch}{1.4}
\setlength{\tabcolsep}{2.5pt}
\begin{table}[htbp]
\caption{Typical property ranges of LCF and HCF magnet at $20^\circ\text{C}$}
\label{tab:materials}
\centering
\footnotesize
\begin{tabular}{|c|c|c|c|c|c|}
\hline
\textbf{Magnet} & \textbf{$iH_c$ (kA/m)} & \textbf{$B_r$ (T)} & \textbf{$\mu_r$} & \textbf{$\mu_g$} & \textbf{Temp. sensitivity} \\
\hline
AlNiCo & 40--130   & 0.6--1.35 & 3--5   & High & Very low              \\
\hline
MnBi             & 100--300  & 0.4--0.6  & 1.1--1.3 & Negligible      & Positive coeff. \\
\hline
MnAl             & 60--120   & 0.5--0.8  & 1.2--1.4 & Low             & Moderate        \\
\hline
FeCrCo           & 20--60    & 1.0--1.4  & 3--5   & Low             & Low              \\
\hline
FeN    & $<$240   &  $>$1.0  & -  & High             & Very low         \\
\hline
NdFeB    & 800--1000 & 1.2-1.5  & 1.05  & Negligible      & Moderate--negative \\
\hline
\end{tabular}
\end{table}

\section{Studied Motor and Simulation Setup}

\subsection{Studied Motor}

\begin{figure}[htbp]
\centering
\includegraphics[width=0.4\textwidth]{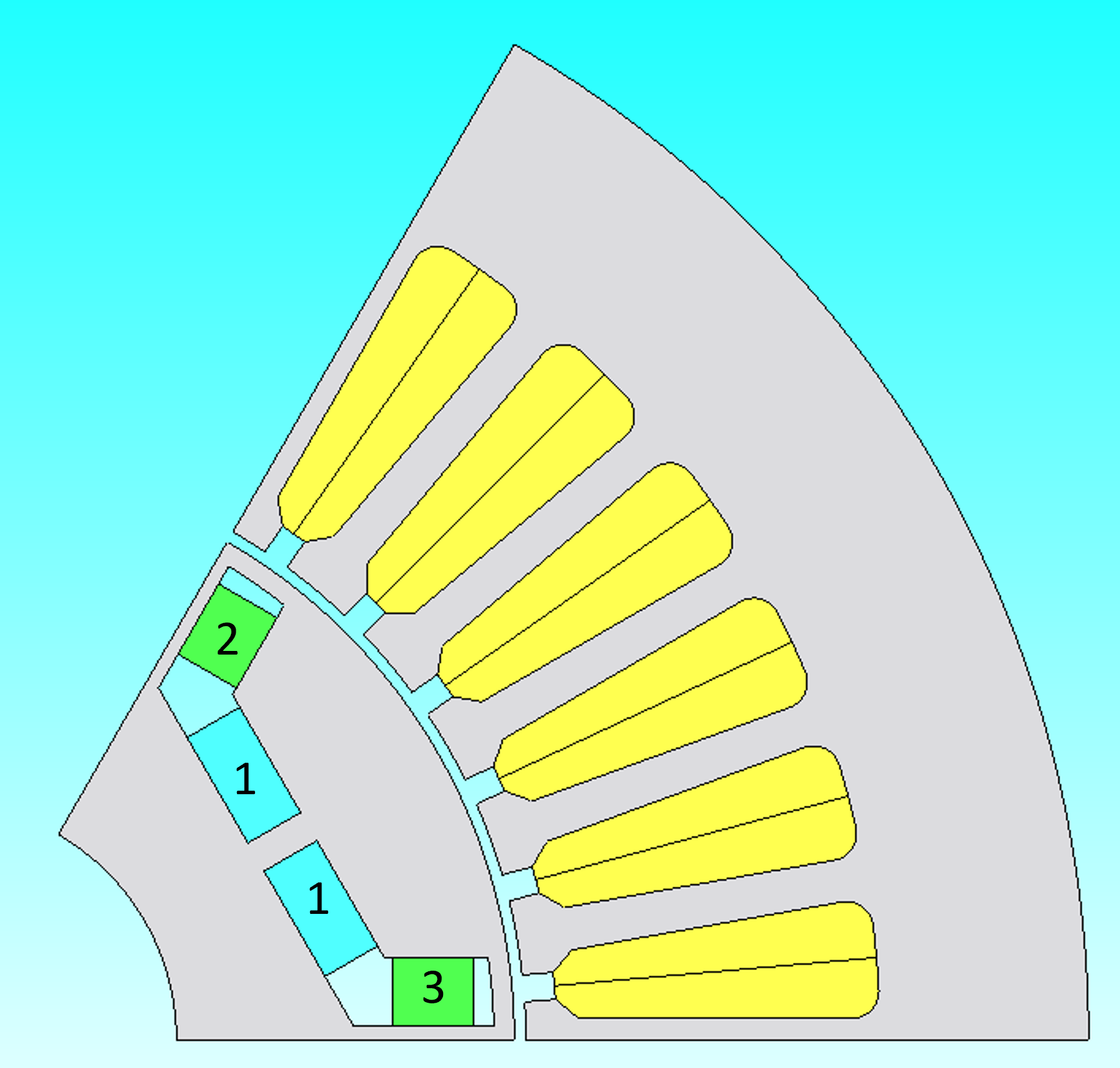}
\caption{Cross-sectional view of the studied motor}
\label{fig:motor}
\end{figure}

The motor under investigation is based on the interior permanent magnet synchronous machine (IPMSM) design documented in the JMAG Application Catalog 255. A cross-sectional view of the motor is shown in Fig.~\ref{fig:motor}. Table~\ref{tab:motor_specs} summarizes the key geometrical and rated parameters of the motor. Each pole incorporates two LCF magnets and two HCF magnets in a U‑shaped arrangement.

\subsection{Simulation Setup}

\begin{figure}[htbp]
\centering
\includegraphics[width=0.95\textwidth]{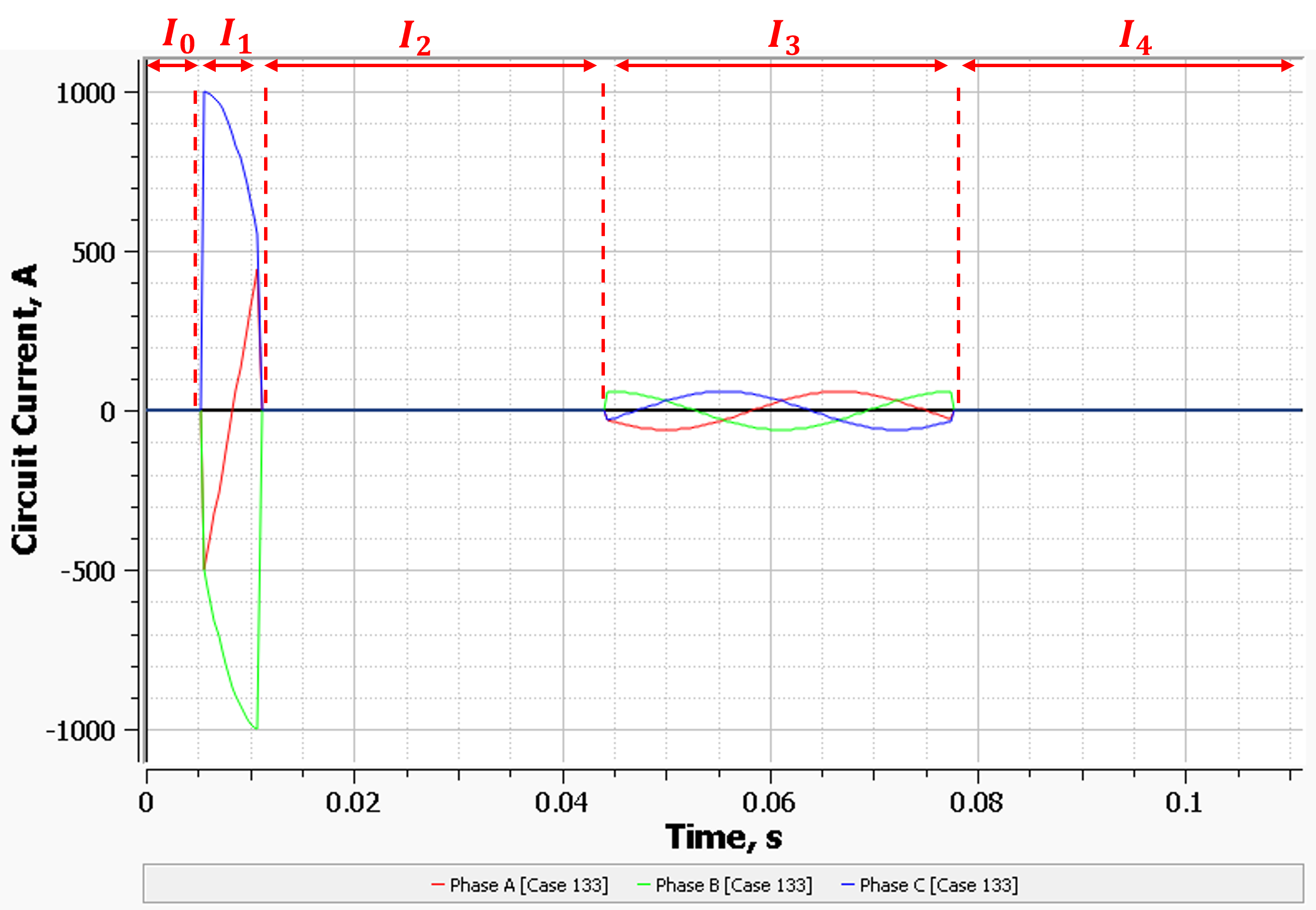}
\caption{Phases current illustrating the five  intervals}
\label{fig:current_waveform}
\end{figure}

To assess the magnetization, demagnetization, and resulting magnetization states of the LCF magnets, a multi‑interval simulation sequence is defined. The initial state of both magnets is set to fully demagnetized ($B_r = 0$ T). The sequence comprises five intervals, as illustrated in the current waveform of Fig.~\ref{fig:current_waveform} and detailed in Table~\ref{tab:intervals}. A full electrical period $T$ corresponds to the fundamental frequency at rated speed. The magnetization pulse is applied during T/6 which is derived to cover the number of slots per phase per pole. The simulations are performed using finite‑element analysis (JMAG) with the motor model and materials defined above.

\begin{table}[htbp]
\caption{Information on the studied motor}
\label{tab:motor_specs}
\centering
\footnotesize
\begin{tabular}{|l|c|}
\hline
\textbf{Parameter} & \textbf{Value} \\
\hline
Number of poles       & 6 \\
\hline
Number of stator slots & 36 \\
\hline
Rated speed (r/min)    & 3000 \\
\hline
Stator outer diameter (mm) & 170 \\
\hline
Air gap length (mm)        & 0.85 \\
\hline
Stack length (mm)          & 100 \\
\hline
Rotor Core          & 50A1000 (JSOL) \\
\hline
Stator Core          & 50A1000 (JSOL) \\
\hline
Magnet 1 (HCF)                  & NdFeB ($B_r=1.2$ T) \\
\hline
Magnets 2-3 (LCF)                  & \makecell{$B_r=1$\,T, $iH_c=110$\,kA/m, \\ $\mu_r=1.1$, $\mu_g=100$, $R=100\,kA/m$} \\
\hline
\end{tabular}
\end{table}

\begin{table}[htbp]
\caption{Intervals of simulation setup (corresponding to Fig.~\ref{fig:current_waveform}).}
\label{tab:intervals}
\centering
\footnotesize
\begin{tabular}{|c|l|c|c|}
\hline
\textbf{Interval} & \textbf{Description} & \textbf{Duration} & \textbf{Applied current} \\
\hline
0 & Initial state ($B_r=0$ T) & $T/6$   & $i_d = 0$, $i_q = 0$ \\
1 & Magnetization pulse         & $T/6$   & $i_d = +1000$ A, $i_q = 0$ \\
2 & No‑load  & $T$      & $i_d = 0$, $i_q = 0$ \\
3 & On‑load operation           & $T$      & $\theta = -90^\circ$ to $90^\circ$, $i = 0$ to $60$\,A \\
4 & No‑load          & $T$      & $i_d = 0$, $i_q = 0$ \\
\hline
\end{tabular}
\end{table}

\section{Magnetization State: Definitions}

To quantitatively evaluate the magnetization state of permanent magnets under various operating conditions, four distinct definitions are introduced: definitions 1 and 2 relate to intrinsic material properties, while definitions 3 and 4 relate to motor-level and control-level quantities.

\begin{itemize}
    \item \textbf{Definition 1:}
\end{itemize}

The first definition utilizes the magnetic flux density \(\vec{B}^{(3)}\) integrated over the magnet volume during on-load operation (interval 3):

\begin{align}
\text{MS(B)} &= \frac{\displaystyle \int_S \left( \vec{B}^{(3)} \cdot \hat{n}_m \right) \, dV}{V_{\text{mag}} \cdot B_r}
\end{align}

where \(\hat{n}_m\) is the unit vector in the direction of magnetization, \(V_{\text{mag}}\) is the total magnet volume, and \(B_r\) is the remanent flux density of the magnet material.

\begin{itemize}
    \item \textbf{Definition 2:}
\end{itemize}

The second definition employs the magnetic polarization \(\vec{J}^{(3)} = \mu_0 \vec{M}^{(3)}\), evaluated during on-load (interval 3):

\begin{align}
\text{MS(J)} &= \frac{\displaystyle \int_S \left( \vec{J}^{(3)} \cdot \hat{n}_m \right) \, dV}{V_{\text{mag}} \cdot J_r}
\end{align}

where \(J_r\) is the remanent polarization. This definition is particularly relevant for assessing irreversible demagnetization risks, as polarization directly reflects the alignment of magnetic domains within the material.

\begin{itemize}
    \item \textbf{Definition 3:}
\end{itemize}

Moving to motor-level quantities, the third definition compares the fundamental component of phase flux linkage under different operating intervals:

\begin{equation}
\text{MS}(\Phi) = cos(\delta) \frac{\Phi_{\text{fund}}^{(4)}}{\Phi_{\text{fund}}^{(2)}}
\end{equation}

where \(\Phi_{\text{fund}}^{(4)}\) represents the fundamental of flux linkage in phase A during operating interval 4, \(\Phi_{\text{fund}}^{(2)}\) is the fundamental of flux linkage during interval 2, and $\delta$ is the shift in angle between the fundamental waveforms of Interval 4 and Interval 2. The ratio indicates the extent of flux reduction (or increase) under load current. The term $cos(\delta)$ accounts for cases where north poles switch to south poles and vice versa due to strong demagnetization. It also accounts for the case where the d axis is shifted due to strong q current.

\begin{itemize}
    \item \textbf{Definition 4:}
\end{itemize}

The fourth definition employs the fundamental component of back electromotive force, which is directly proportional to flux linkage in the absence of significant saturation effects:

\begin{equation}
\text{MS}(E) = cos(\delta) \frac{E_{\text{fund}}^{(4)}}{E_{\text{fund}}^{(2)}}
\end{equation}

where \(E_{\text{fund}}^{(4)}\) and \(E_{\text{fund}}^{(2)}\) are the fundamental components of back EMF in phase A during intervals 4 and 2, respectively, and $\delta$ is the shift in angle between the fundamental waveforms of Interval 4 and Interval 2. Back EMF is readily measurable and directly influences the voltage requirements of the drive system. The term $cos(\delta)$ accounts for the same effects mentioned previously.

The first two definitions are material-level and are typically evaluated through finite element analysis by integrating over magnet volumes or surfaces, while the latter two are motor-level and can be assessed through both simulation and experimental measurements, making them valuable for validation and control algorithm development.

\section{Results}

This section presents the results of the magnetization state definitions introduced in Section IV. Each definition is evaluated across the \(i_d, i_q\) operating plane.

\begin{figure}[htbp]
\centering

\begin{subfigure}[b]{0.48\textwidth}
    \centering
    \includegraphics[width=\textwidth]{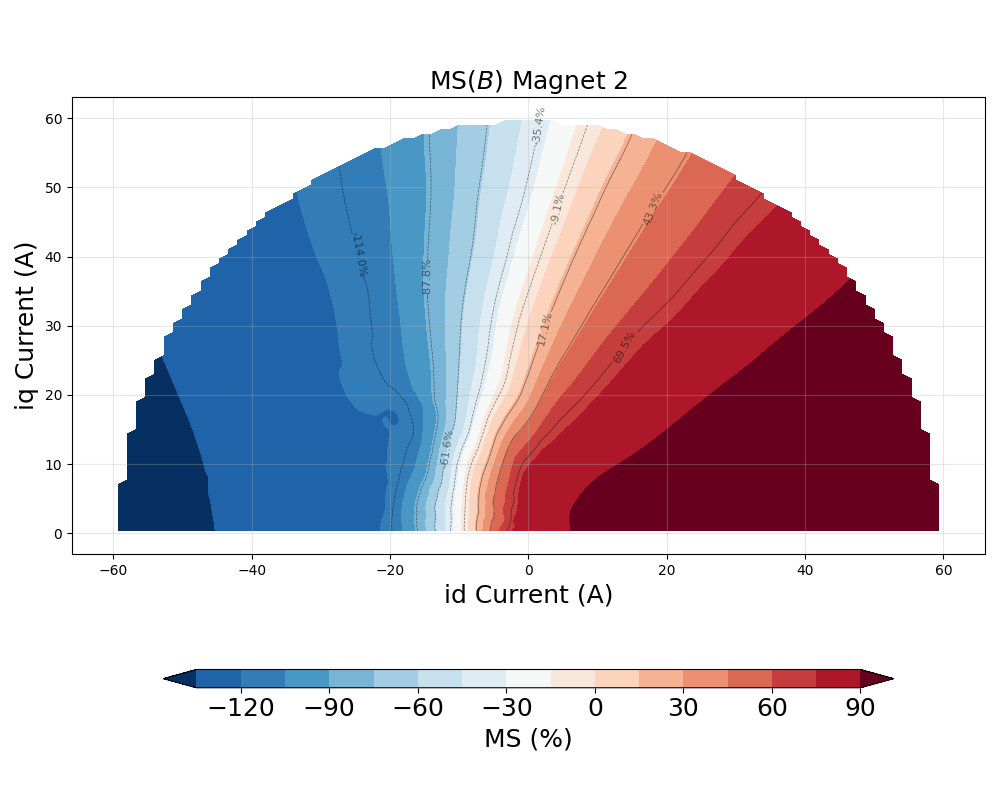}
    \caption{Magnet 2}
    \label{fig:ms1_pm2}
\end{subfigure}
\hfill
\begin{subfigure}[b]{0.48\textwidth}
    \centering
    \includegraphics[width=\textwidth]{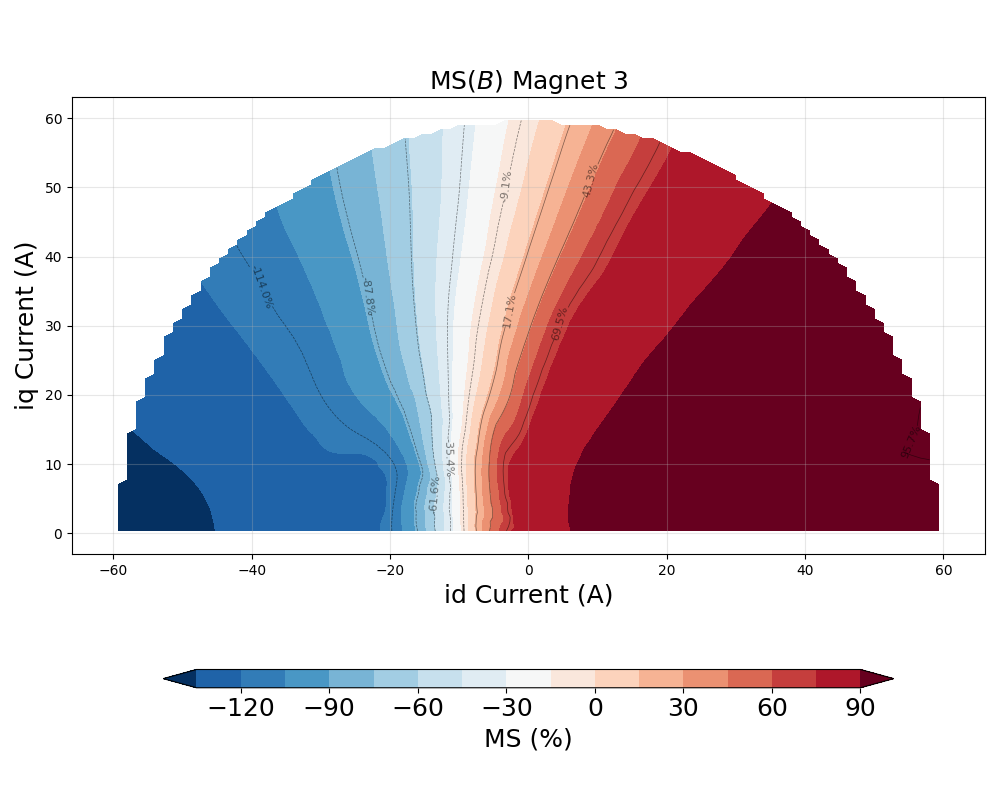}
    \caption{Magnet 3}
    \label{fig:ms1_pm3}
\end{subfigure}

\caption{Magnetization state MS(B) according to Definition 1 for both magnets}
\label{fig:ms1_both}
\end{figure}

Figures \ref{fig:ms1_pm2} and \ref{fig:ms1_pm3} show the magnetization state according to definition 1 (MS(B)) for both magnets 2 and 3, respectively. It can be observed that Magnet 3 shows an MS above 90\% for a wider range in $i_d, i_q$ plane. This difference is because a positive $i_q$ current will magnetize magnet 3 and demagnetize magnet 2. Further, negative values of MS show that the magnets have a reverse magnetization direction due to negative $i_d$ values applied during the Interval 3.

\begin{figure}[htbp]
\centering

\begin{subfigure}[b]{0.48\textwidth}
    \centering
    \includegraphics[width=\textwidth]{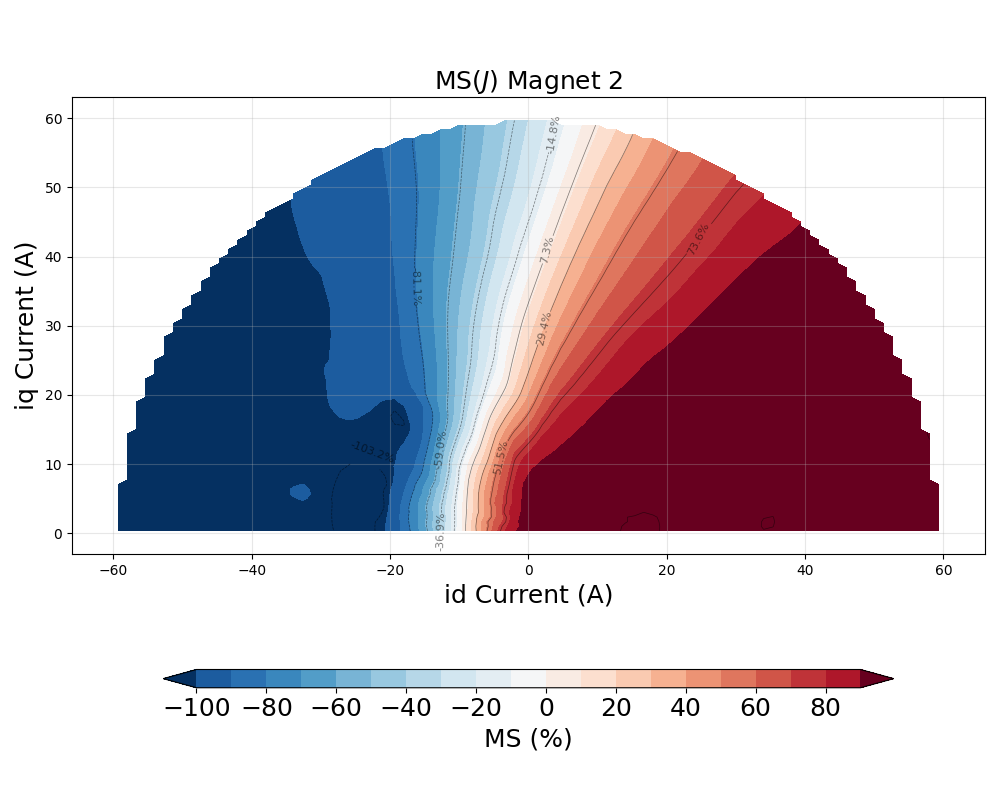}
    \caption{Magnet 2}
    \label{fig:ms2_pm2}
\end{subfigure}
\hfill
\begin{subfigure}[b]{0.48\textwidth}
    \centering
    \includegraphics[width=\textwidth]{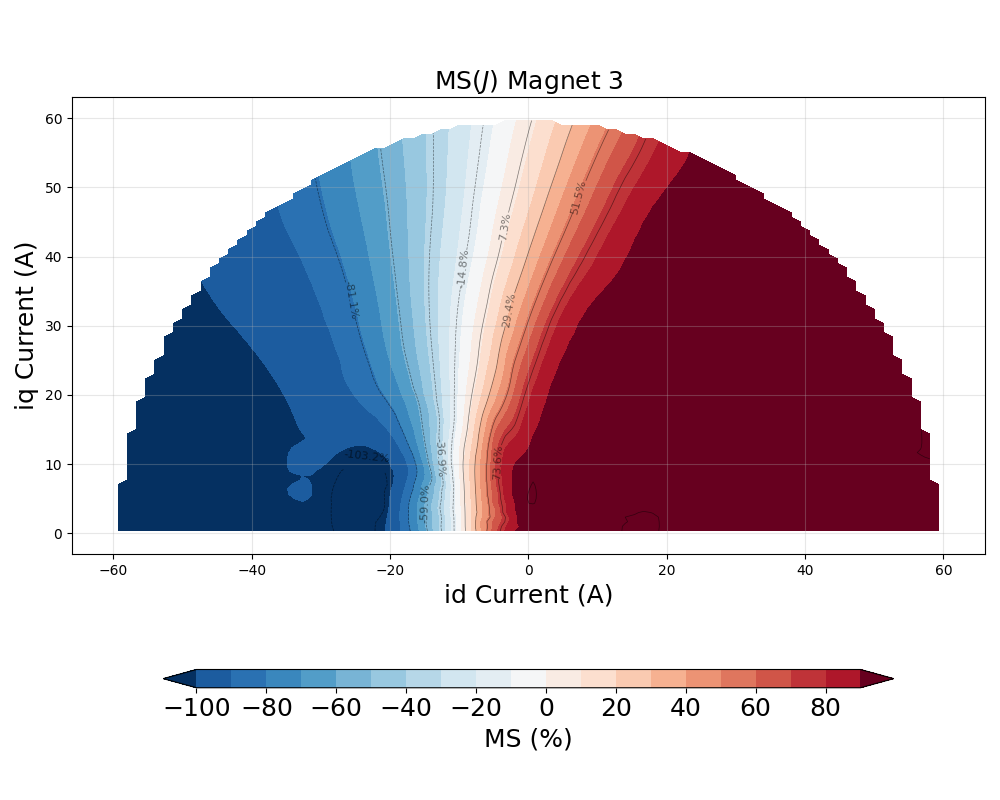}
    \caption{Magnet 3}
    \label{fig:ms2_pm3}
\end{subfigure}

\caption{Magnetization state MS(J) according to Definition 2 for both magnets}
\label{fig:ms2_both}
\end{figure}

Figures \ref{fig:ms2_pm2} and \ref{fig:ms2_pm3} show the magnetization state according to definition 2 (MS(J)) for both magnets 2 and 3, respectively. MS using polarization $J$ gives slightly different results compared to B-based definition. We still notice that magnet 3 has a wider range of high MS value for the same reasons mentioned previously. This MS definition is more suitable for characterizing the magnetization of the magnet as it is directly related to its inherent property, i.e. polarization.

\vspace{-2pt}

\begin{figure}[htbp]
\centering

\begin{subfigure}[b]{0.48\textwidth}
    \centering
    \includegraphics[width=\textwidth]{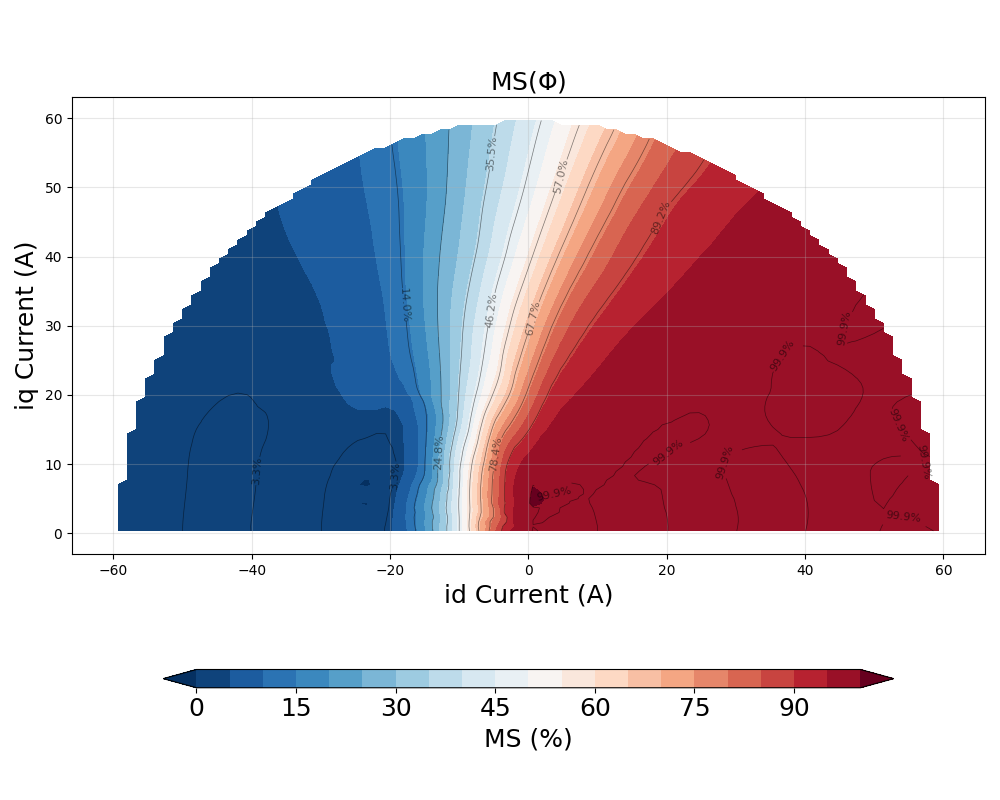}
    \caption{MS($\Phi$)}
    \label{fig:ms3}
\end{subfigure}
\hfill
\begin{subfigure}[b]{0.48\textwidth}
    \centering
    \includegraphics[width=\textwidth]{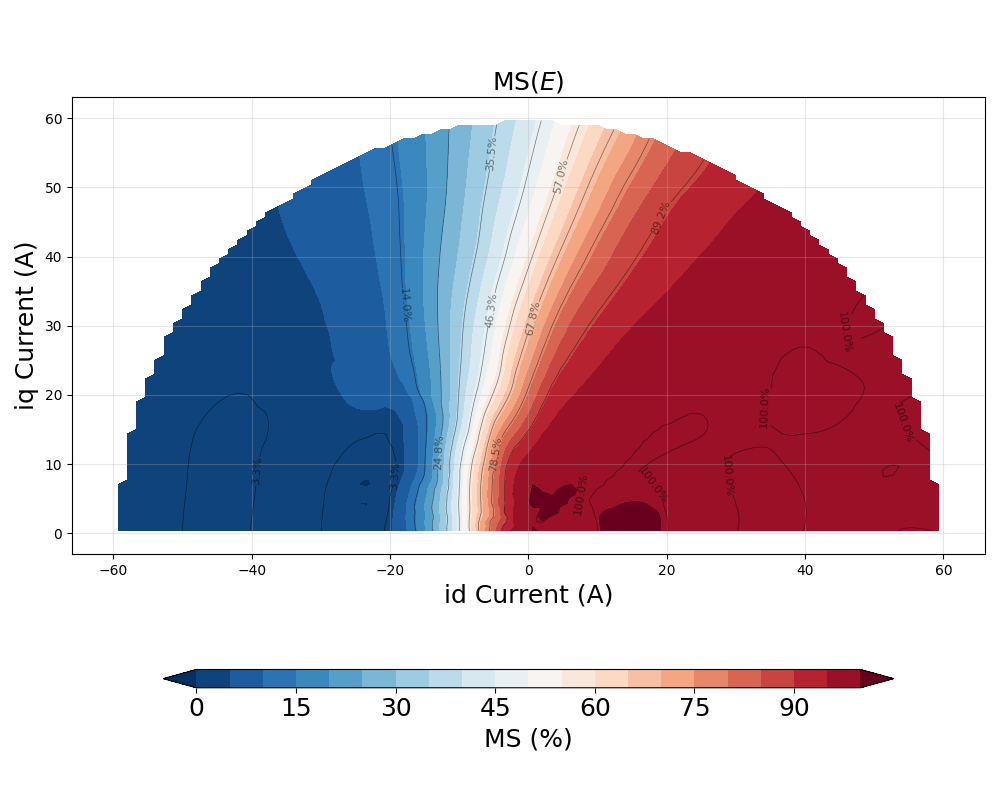}
    \caption{MS($E$)}
    \label{fig:ms4}
\end{subfigure}

\caption{Magnetization state according to (a) Definition 3: MS($\Phi$) and (b) Definition 4: MS($E$)}
\label{fig:ms3_4}
\end{figure}

Figures \ref{fig:ms3} and \ref{fig:ms4} present the magnetization state according to definition 3 (MS($\Phi$)) and definition 4 (MS(E)), respectively. Both definitions based on flux and back-emf yield almost similar results. Although magnets undergo demagnetization leading to reverse magnetization direction when applying the load current (interval 3), poles are not reversed when removing the load current (interval 4) as noticed by positive values of MS for both flux and back-emf-based definitions and this is mainly due to HCF magnets that still dictate flux direction. These two definitions are more suitable for motor control.

\vspace{6pt}

\section*{Conclusion}

This paper has proposed and compared four definitions for quantifying the magnetization state of Low Coercive Force magnets in variable flux memory motors. The definitions span from intrinsic material properties, based on magnetic flux density B and magnetic polarization J, to motor-level metrics derived from fundamental flux linkage and back-EMF components. Through finite element analysis across the \(i_d, i_q\) operating plane, the magnetization state has been examined using the four definitions. Definitions based on flux linkage and back EMF operate at the motor level. While they yield results that are consistent with each other, they differ from those obtained through material-based definitions (such as B and J). The material-based approaches remain essential for design validation and demagnetization risk assessment, whereas the motor-level metrics offer practical advantages for real-time estimation and control applications.


\bibliographystyle{unsrtnat}
\bibliography{refs}

\end{document}